\documentclass[conference]{IEEEtran}

\usepackage[utf8]{inputenc}
\usepackage{amsmath, commath}
\usepackage{amsthm}
\usepackage{amsfonts}
\usepackage{amssymb}
\usepackage{bm}
\usepackage{bbm}
\usepackage{graphicx}
\usepackage{graphics}
\usepackage{float}
\usepackage{tikz}
\usetikzlibrary{shapes, snakes, patterns}
\makeatletter
\newcommand*{\rom}[1]{\expandafter\@slowromancap\romannumeral #1@}

\makeatother
\usepackage{epstopdf}
\usepackage[export]{adjustbox}
\usepackage[list=true]{subcaption}
\usepackage{color}
\usepackage{algorithm}
\usepackage{algorithmic}

\usepackage[justification=centering]{caption}
\usepackage{enumerate} 
\usepackage{cite}
\usepackage{suffix}
\usepackage{mathtools}
\usepackage{tabularx, booktabs}
\usepackage{multirow}
\usepackage{diagbox}

\def\x{{\mathbf x}}

\def\h{{\mathbf h}}

\def\y{{\mathbf y}}

\def\n{{\mathbf n}}
\def\x{{\mathbf x}}
\def\0{{\mathbf 0}}

\def\i1{{\mathbf 1}}

\usepackage{dsfont} 

\DeclarePairedDelimiter\ceil{\lceil}{\rceil}

\title{On the Use of CSI for the Generation of RF Fingerprints and Secret Keys}
\author{Muralikrishnan Srinivasan, Sotiris Skaperas and Arsenia Chorti\\
ETIS UMR 8051 / CY Paris University, ENSEA, CNRS

\thanks{\textbf{This article has been accepted for publication in The 25th International ITG Workshop on Smart Antennas (WSA 2021)}}
}

\IEEEoverridecommandlockouts

\begin{document}

\maketitle

\begin{abstract}
    This paper presents a systematic approach to use channel state information for authentication and secret key distillation for physical layer security (PLS). We use popular machine learning (ML) methods and signal processing-based approaches to disentangle the large scale fading and be used as a source of uniqueness, from the small scale fading, to be treated as a source of shared entropy secret key generation (SKG). The ML-based approaches are completely unsupervised and hence avoid exhaustive measurement campaigns. We also propose using the Hilbert Schmidt independence criterion (HSIC); our simulation results demonstrate that the extracted stochastic part of the channel state information (CSI) vectors are statistically independent.
\end{abstract}

\section{Introduction}
 The renewed interest in physical layer security (PLS) technologies for sixth-generation (6G) systems stems from the emergence of massive-scale Internet of things (IoT) networks, which have an extensive range of non-functional (security) constraints as well as computational, power and energy limitations, delay and latency constraints, etc. \cite{zou2016survey, al2015internet}. One of the most popular physical layer security (PLS) techniques is for the transmitter (Alice) and the receiver (Bob) to extract a key from the wireless channel realisations exploiting the common randomness of the wireless channels during the channel coherence time \cite{chorti2016physical, MahdiBookChapter}. 
 
 Wireless channels consist of two parts, namely the large-scale fading, which includes the path loss and shadowing, and a small-scale multipath fading \cite{goldsmith2005wireless}. The path loss is deterministic and therefore is not secure for key generation \cite{zhang2019key}. Shadowing limits the key generation performance due to its slow temporal variation\cite{zhang2020new}. However, the small-scale fading is usually unpredictable and is a valuable source of key generation. In the 6G era of massive and critical IoT, lightweight PLS schemes are now considered. In this contribution, we demonstrate that the wireless communication medium offers the platform for providing building blocks for two cornerstone security operations: 
\begin{enumerate}
\item Authentication through RF fingerprinting. In particular, \textit{large scale fading} that is largely predictable can be treated as a source of \textit{uniqueness} (fingerprint). Importantly, large scale fading effects (path loss and shadowing) are directly related to positioning, allowing for localisation to be used as a second factor of (soft) authentication in multi-factor authentication protocols;
\item Symmetric secret key generation (SKG), i.e., unpredictable variations in the channel state information (CSI) due to \textit{small scale fading} in the wireless channel, that can be treated as a source of \textit{entropy} to distil keys from shared area. We note in passing that in standard cryptography, the concept of "unpredictability'' is central to evaluating the quality of pseudo-random number generators. In essence, the quality of a pseudo-random source is evaluated against the ability to build next bit predictors by a powerful adversary having access to all effective algorithms (i.e., algorithms that can run in polynomial time). 
\end{enumerate}

A secure key generation depends on three principles: channel reciprocity between Alice and Bob, spatial decorrelation and temporal variations \cite{zhang2020new}. Spatial decorrelation is particularly important because a passive eavesdropper (Eve) present close to the legitimate users can generate the duplicate keys by exploiting the shared spatial correlation. Based on Jakes' model, the channel will be uncorrelated when a third party is located half-wavelength away \cite{goldsmith2005wireless}.  

Under this assumption, to facilitate reconciliation, the authors of \cite{li2018high} carry out a theoretical study on pre-processing algorithms such as principal component analysis (PCA) to establish a high-agreement uncorrelated secret key by retaining only the first few dominant components of the channel vectors. The core hypothesis in this work is that the PCA can improve the cross-correlation between the channel measurements of Alice and Bob because the noisy observation is removed by only keeping principal components.

However, experimental results show that a half-wavelength distance spatial decorrelation is valid only in rich scattering environments \cite{edman2016security, zenger2016passive, dautov2019effects, ji2020vulnerabilities}. As a counter-example, in line-of-sight conditions, retaining the first few dominant principal components is equivalent to retaining the eigenvectors related to the predictable large-scale features; therefore, such an approach would be counterproductive in these conditions.

The second important point is that in literature, statistical tests evolve around correlation measures. There is no guarantee that spatial or temporal decorrelation implies independence.  Note that in most works of the existing literature, SKG is performed without systematically removing the predictable spatially or temporally correlated component of the wireless channel coefficients \cite{peng2017secret, li2018high, mitev2020authenticated}. 

A secure key generation process is instrumental for confidentiality and integrity, e.g. when used with symmetric key encryption algorithms. However, secret key generation cannot as such be used for the authentication \cite{zhang2020new}; authentication requires a predictable and verifiable source of uniqueness, such as the node location and RF fingerprinting. In \cite{wang2016physical, li2017fingerprints, fang2018learning}  physical layer authentication approaches are proposed by exploiting different channel parameters. To the best of our knowledge, only a few papers such as \cite{shi2013ask, shi2015mask} aim to achieve both device authentication and SKG simultaneously in the context of body area networks.

Despite the immense bibliography in both RF fingerprinting and SKG, a systematic treatment of the CSI as both a source of uniqueness and a source of entropy is missing. Therefore, in this contribution, we aim at filling this gap and presenting machine learning (ML) and signal processing based approaches to disentangle the large scale fading (source of uniqueness) from the small scale fading (source of shared entropy), inspired by the popular channel charting methods introduced in \cite{studer2018channel, deng2018multipoint, ferrand2020triplet, huang2019improving, lei2019siamese}. Note that these methods are entirely online and do not require any training samples. To ensure secrecy, we not only enforce spatial decorrelation between locations but also guarantee a much stronger spatial independence using the d- variable Hilbert-Schmidt independence criterion (dHSIC) \cite{dhsic}.

 \begin{table*}[t]
 \renewcommand{\arraystretch}{2}
\setlength{\tabcolsep}{14pt}
 \caption{PCA: Mean Correlation coefficient across across the 9-nearest neighbours of every location.}
\setlength{\tabcolsep}{3pt}
\centering 
\label{Table:Corr}
\begin{tabular}{|l|c|c|c|c|c|c|c|c|c|c|c|c|c|c|c|}
\hline
\multicolumn{1}{|c|}{SNR} &  \multicolumn{5}{|c|}{$10~dB$} &  \multicolumn{5}{|c|}{$30~dB$}&  \multicolumn{5}{|c|}{$50~dB$}\\
\hline
\multicolumn{1}{|c|}{SCS ($~KHz$)} & 
\multicolumn{1}{|c|}{ $15$} & \multicolumn{1}{|c|}{$30$} & \multicolumn{1}{|c|}{$60$}
& \multicolumn{1}{|c|}{$100$} & \multicolumn{1}{|c|}{$1000$} & 
\multicolumn{1}{|c|}{ $15$} & \multicolumn{1}{|c|}{$30$} & \multicolumn{1}{|c|}{$60$}
& \multicolumn{1}{|c|}{$100$} & \multicolumn{1}{|c|}{$1000$} & 
\multicolumn{1}{|c|}{ $15$} & \multicolumn{1}{|c|}{$30$} & \multicolumn{1}{|c|}{$60$}
& \multicolumn{1}{|c|}{$100$} & \multicolumn{1}{|c|}{$1000$}\\  
\hline
 Observed channel & $0.25$ & $0.25$ & $0.25$  & $0.31$ & $0.59$ &  $0.34$ & $0.61$ & $0.65$ & $0.88$ & $0.74$ & $0.9$ & $0.97$ & $0.98$ & $0.98$ & $0.74$\\
\hline
 $\hat{D}=1$ & $0.23$ & $0.24$ & $0.24$  & $0.24$ & $0.27$ & $0.23$ & $0.32$ & $0.25$ &  $0.51$ & $0.51$ & $0.52$ & $0.85$ & $0.85$ & $0.95$ & $0.54$ \\
\hline
$\hat{D}=2$ & $0.24$ & $0.23$ & $0.23$  & $0.24$ & $0.25$ & $0.24$ & $0.24$ & $0.24$ & $0.24$ & $0.51$ & $0.24$& $0.26$ & $0.24$ & $0.46$ & $0.57$\\
\hline
 $\hat{D}=3$ & $0.24$ & $0.23$ & $0.23$ & $0.24$ & $0.24$ & $0.24$ & $0.24$ & $0.24$ & $0.24$ & $0.42$ & $0.24$& $0.24$ & $0.23$ & $0.24$ & $0.59$\\
\hline
$\hat{D}=4$ & $0.24$ & $0.24$ & $0.23$ & $0.23$ & $0.24$ & $0.24$ & $0.23$ & $0.24$ & $0.24$ & $0.31$ & $0.24$ & $0.24$ & $0.24$ & $0.24$ & $0.58$\\
\hline
$\hat{D}=8$ & $0.24$ & $0.24$ & $0.24$ & $0.24$ & $0.24$ & $0.24$ & $0.24$ & $0.24$ & $0.24$ & $0.24$ & $0.24$ & $0.24$ & $0.23$ & $0.24$ & $0.26$ \\
\hline
\end{tabular}
\end{table*}  

\section{Pre-processing Using PCA}

Consider single-antenna legitimate nodes, referred to as Alices and a base station referred to as Bob, over a multicarrier fading channel. Alices' spatial locations are denoted by $\{\x_n\}_{n=1}^{N}$ $n= 1,\hdots,N$, where $\{x_n\}_{n=1}^{N} \in \mathbb{R}^L$ and $L$ denotes the spatial dimensions considered (typically $L=2$). Let the channel function mapping the spatial locations to the $M\times{1}$ CSI vectors $\{\h_n\}_{n=1}^{N}$ denoted by $\mathcal{H} : \mathbb{R}^L \to \mathbb{C}^M$, where $M$ is the number of orthogonal frequency division multiplexing (OFDM) subcarriers.
Alice and Bob exchange pilot signals so that their respective observations can be modelled as
\begin{equation}
    \y_{nu} = \h_n x +\n_{nu}, \: n=1, \hdots, N \text{, } u\in\{a, b\},
\end{equation}
where the index $a$ denotes Alice, $b$ denotes Bob; $\n_{na}$ and $\n_{nb}$ are complex circularly symmetric Gaussian noise variables and the pilot symbols $x$ are chosen from binary phase-shift keying (BPSK) constellation. The channel estimates at Alice and Bob, respectively, are denoted by $\hat \h_{na}= \y_{na}$ and  $\hat \h_{nb}= \y_{nb}$ for $ n=1, \hdots, N$.

Inspired from the popular channel charting methods introduced in \cite{studer2018channel}, we learn the functional mapping that captures the dominant predictable spatially correlated components of the CSI vectors. 
Among the various approaches like principal component analysis (PCA), autoencoders, etc., for performing channel-charting, in this paper, we focus primarily on PCA to determine the first $\hat{D}$ dominant components from each of the $M$ dimensional channel vector \cite{studer2018channel}. Let $\hat{\bm{H}}_u= \left[\hat {\bm{h}}_{1u}, \cdots, \hat {\bm{h}}_{Nu} \right]$ denote the observed channel. PCA transforms the observed  $M \times N$ matrix $\hat{\bm{H}}_u$ into the lower dimension, $\widehat{D}$-dimensional channel matrix $\hat {\bm{W}}_{u}$, where $\widehat{D}$ is the new number of variables and $\widehat{D}<M$. The PCA transformation is given by the $M\times{\widehat{D}}$ matrix $\bm{U}$, such that,   
\begin{equation}
    \hat {\bm{W}}_{u} = \hat{\bm{H}}_u \bm{U}_{\widehat{D}}
\end{equation}
where $\bm U_{\widehat{D}}$ is the matrix whose columns are the eigenvectors of the matrix Cov$\left(\widehat{\mathbf{H}}_u\right)$, corresponding to the dominant $\widehat{D}$ eigenvalues.

The inverse PCA gives us the reconstructed form of the observed channel, based on the first $\widehat{D}$ principal components (PCs) as follows,
\begin{equation}
    \tilde{\bm{H}}_{u} = \hat {\bm{W}}_{u}\bm U_{\widehat{D}}',
\end{equation}
where $ \tilde{\bm{H}}_{u} =\left[\tilde {\bm{h}}_{1u}, \cdots, \tilde {\bm{h}}_{Nu} \right]$ for $u\in\{a, b\}$ is a $N\times{M}$ matrix.
Finally, the residuals of the inverse PCA are given by, 
\begin{equation}
\left \{\hat{\bm{z}}_{nu}(\hat {D})\right \}_{n=1}^{N}= \{\hat{\bm{h}}_{nu}-\tilde{\bm{h}}_{nu}\}_{n=1}^{N}, \text{ for} \: u\in\{a, b\}.
\end{equation}
where $\{\tilde{\bm{h}}_{nu}\}_{n=1}^{N}$, considering the structure of (4), can also be called as the predictable part of the observed channel.

To efficiently decompose (4) in terms of the predictable and the unpredictable part, the number of PCs $\widehat{D}$ have to be chosen such that the residuals are uncorrelated and independent. In the following subsections we evaluate the spatial correlation and independence of the residuals $\left \{\hat{\bm{z}}_{nu}(\hat {D})\right \}_{n=1}^{N}$ by means of Pearson correlation coefficient and $dHSIC$. If the residuals are uncorrelated, or more strongly, if they are independent, implying that their behaviour is also unpredictable, they can be used for SKG and withstand passive eavesdropping attacks.

\subsection{Correlation Coefficient}
The straightforward metric to measure the degree of spatial decorrelation of the residuals between locations is the Pearson correlation coefficient. Given a pair of residuals $\hat{\bm{z}}_{n_1u}$ and $\hat{\bm{z}}_{n_2u}$ at two locations $n_1$ and $n_2$ respectively, the Pearson correlation coefficient is given by
\begin{equation}
    \rho_u(n_1, n_2)= \frac{\mathbb{E}\left(\hat{\bm{z}}_{n_1u}-\mathbb E\left( \hat{\bm{z}}_{n_1u}\right)\right) \mathbb{E}\left(\hat{\bm{z}}_{n_2u}-\mathbb E\left( \hat{\bm{z}}_{n_2u}\right)\right)}{\sigma_{n_1u} \sigma_{n_2u}},
\end{equation}
where $\sigma_{n_1u}$ and $\sigma_{n_2u}$ are the respective standard deviations.

\begin{table*}[t]
 \renewcommand{\arraystretch}{2}
\setlength{\tabcolsep}{14pt}
 \caption{Rejection rate of the null hypothesis of independence applying the $dHSIC$ across the 9-nearest neighbours of every location.}
\setlength{\tabcolsep}{3pt}
\centering 
\label{Table:Ind}

\begin{tabular}{|l|c|c|c|c|c|c|c|c|c|c|c|c|c|c|c|}
\hline
\multicolumn{1}{|c|}{SNR} &  \multicolumn{5}{|c|}{$10~dB$} &  \multicolumn{5}{|c|}{$30~dB$}&  \multicolumn{5}{|c|}{$50~dB$}\\
\hline
\multicolumn{1}{|c|}{SCS ($~KHz$)} & 
\multicolumn{1}{|c|}{ $15$} & \multicolumn{1}{|c|}{$30$} & \multicolumn{1}{|c|}{$60$}
& \multicolumn{1}{|c|}{$100$} & \multicolumn{1}{|c|}{$1000$} & 
\multicolumn{1}{|c|}{ $15$} & \multicolumn{1}{|c|}{$30$} & \multicolumn{1}{|c|}{$60$}
& \multicolumn{1}{|c|}{$100$} & \multicolumn{1}{|c|}{$1000$} & 
\multicolumn{1}{|c|}{ $15$} & \multicolumn{1}{|c|}{$30$} & \multicolumn{1}{|c|}{$60$}
& \multicolumn{1}{|c|}{$100$} & \multicolumn{1}{|c|}{$1000$}\\  
\hline
Observed channel & $0.05
$ & $0.04$  & $0.08$ & $0.03$ & $0.41$ &  $0.05$ & $0.41$ & $0.13$ & $0.88$ & $1$ & $0.72$ & $1$ & $1$ & $1$ & $1$\\
\hline
 $\hat{D}=1$ & $0.05$ & $0.04$  & $0.05$ & $0.05$ & $0.18$ & $0.05$ & $0.3$ & $0.06$ &  $0.86$ & $1$ & $0.92$ & $0.98$ & $0.98$ & $1$ & $1$ \\
\hline
$\hat{D}=2$ & $0.02$ & $0.06$ & $0.04$  & $0.05$ & $0.05$ & $0.04$ & $0.04$ & $0.03$ & $0.07$ & $0.98$ & $0.05$& $0.06$ & $0.04$ & $0.56$ & $1$\\
\hline
 $\hat{D}=3$ & $0.04$ & $0.05$ & $0.04$ & $0.05$ & $0.05$ & $0.04$ & $0.04$ & $0.04$ & $0.07$ & $0.59$ & $0.04$& $0.04$ & $0.05$ & $0.04$ & $1$\\
\hline
 $\hat{D}=4$ & $0.06$ & $0.04$ & $0.05$ & $0.04$ & $0.06$ & $0.06$ & $0.05$ & $0.05$ & $0.04$ & $0.16$ & $0.02$& $0.04$ & $0.04$ & $0.06$ & $1$\\
\hline
$\hat{D}=8$ & $0.05$ & $0.06$ & $0.05$ & $0.04$ & $0.04$ & $0.04$ & $0.03$ & $0.04$ & $0.06$ & $0.09$ & $0.03$ & $0.04$ & $0.05$ & $0.04$ & $0.1$ \\
\hline
\end{tabular}
\end{table*}

\begin{figure}
    \centering
    \begin{subfigure}[b]{0.25\textwidth}
    \includegraphics[width=\textwidth]{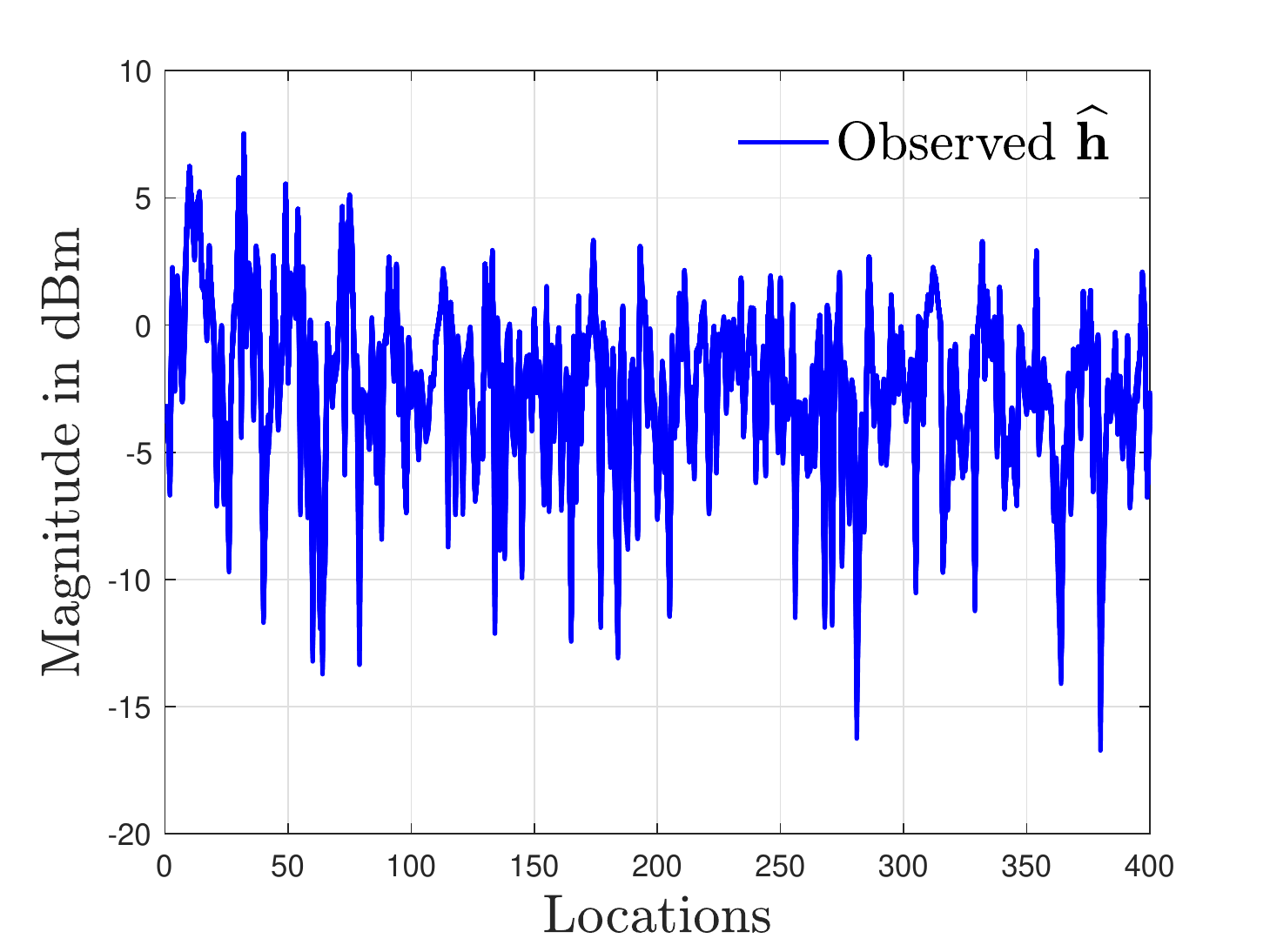}
    \caption{}
    \end{subfigure}%
    \begin{subfigure}[b]{0.25\textwidth}
    \includegraphics[width=\textwidth]{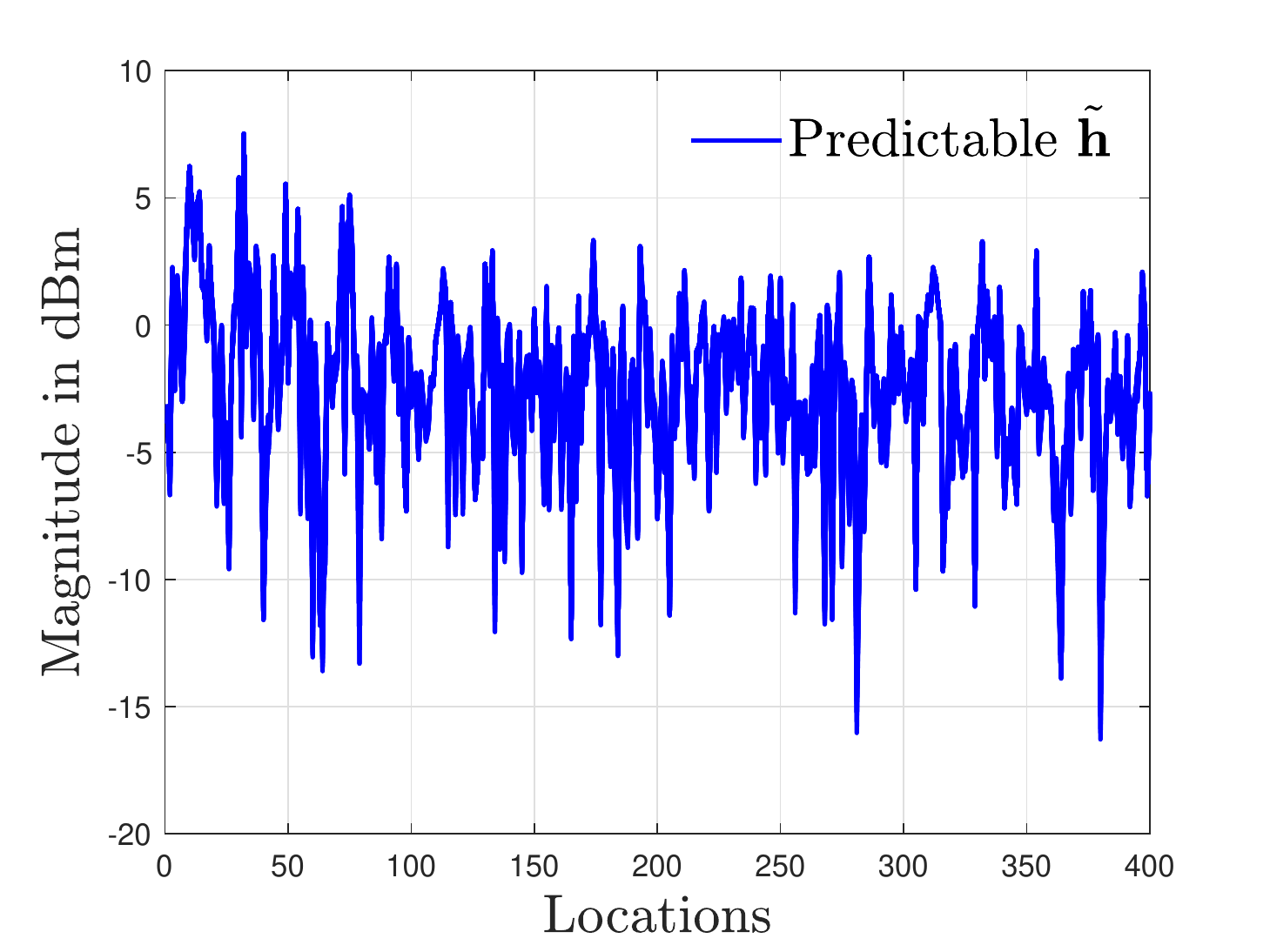}
    \caption{}
    \end{subfigure}
    \begin{subfigure}[b]{0.25\textwidth}
    \includegraphics[width=\textwidth]{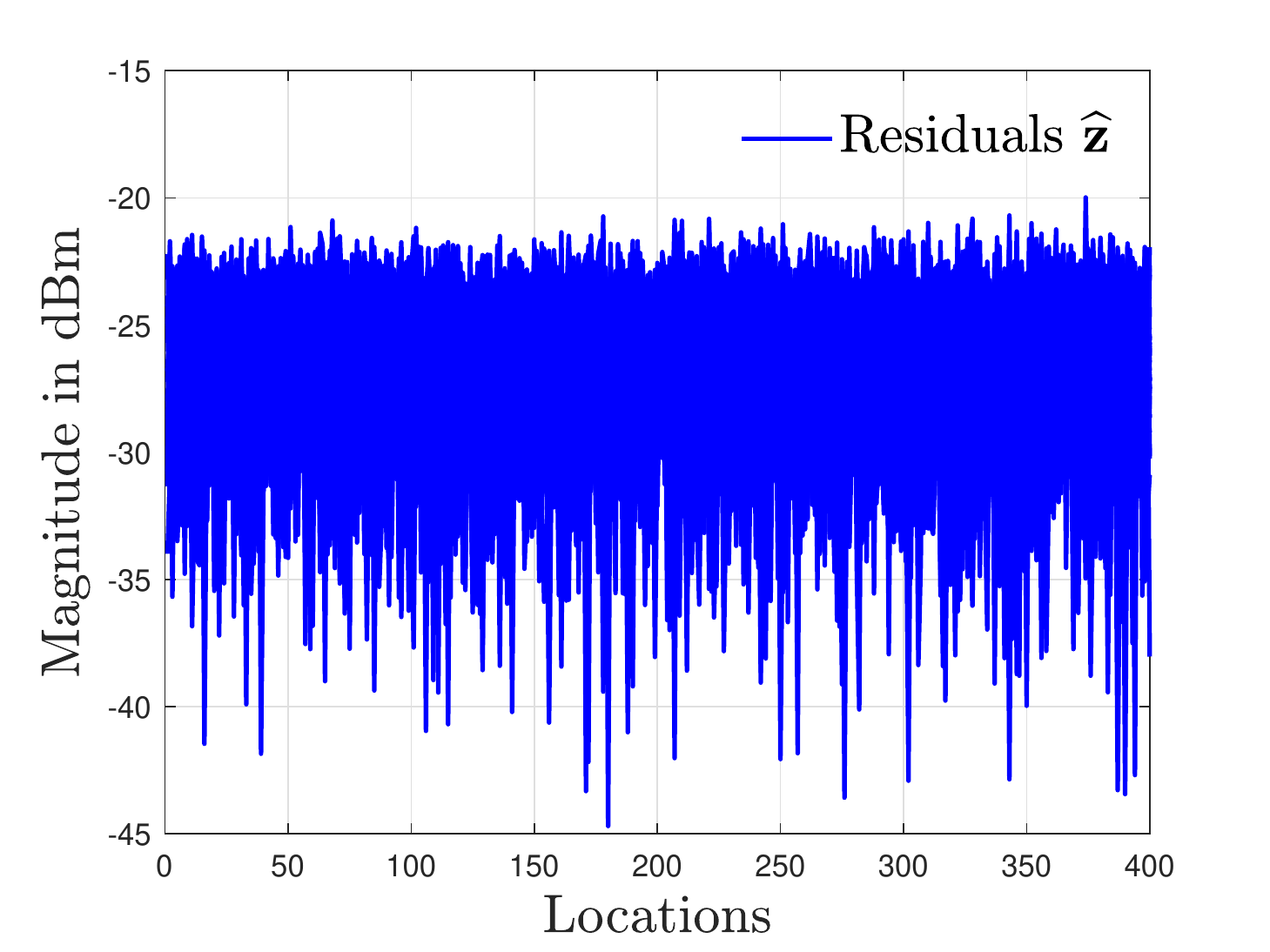}
    \caption{}
    \end{subfigure}%
    \caption{The a) observed, b) predictable and c) residuals part of the CSI, for $SCS = 100$ KHz, $SNR = 50$ dB and $\widehat{D}=3$.}
    \label{fig: OPR}
\end{figure} 

\begin{figure*}[t]
     \centering
     \begin{subfigure}{0.32\textwidth}
         \centering
         \includegraphics[width=\textwidth]{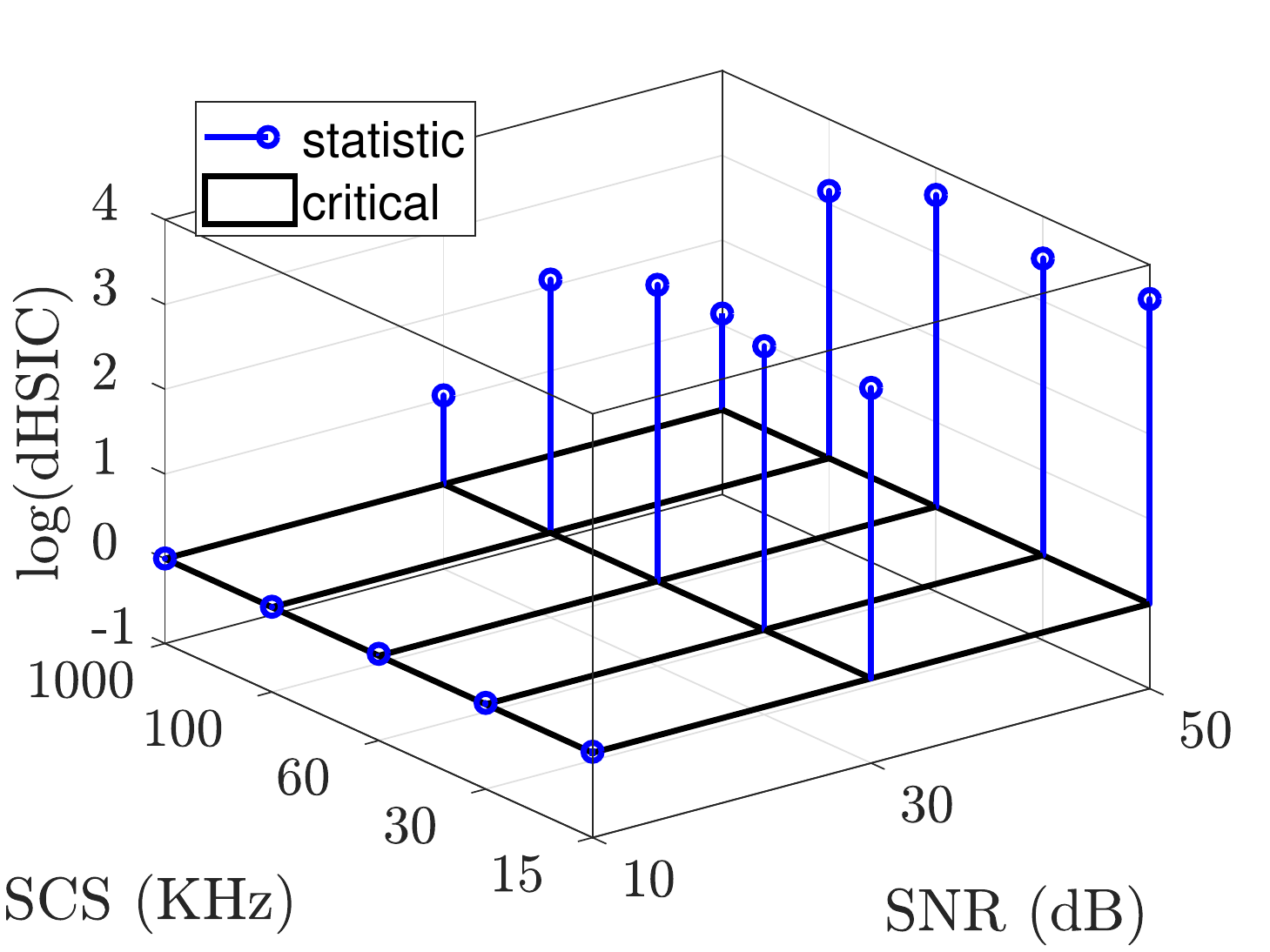}
         \caption{Observed channel ${\widehat{\bf{h}}}$}
         \label{fig:SCC15dB0dim}
     \end{subfigure}
     \hfill
     \begin{subfigure}{0.32\textwidth}
         \centering
         \includegraphics[width=\textwidth]{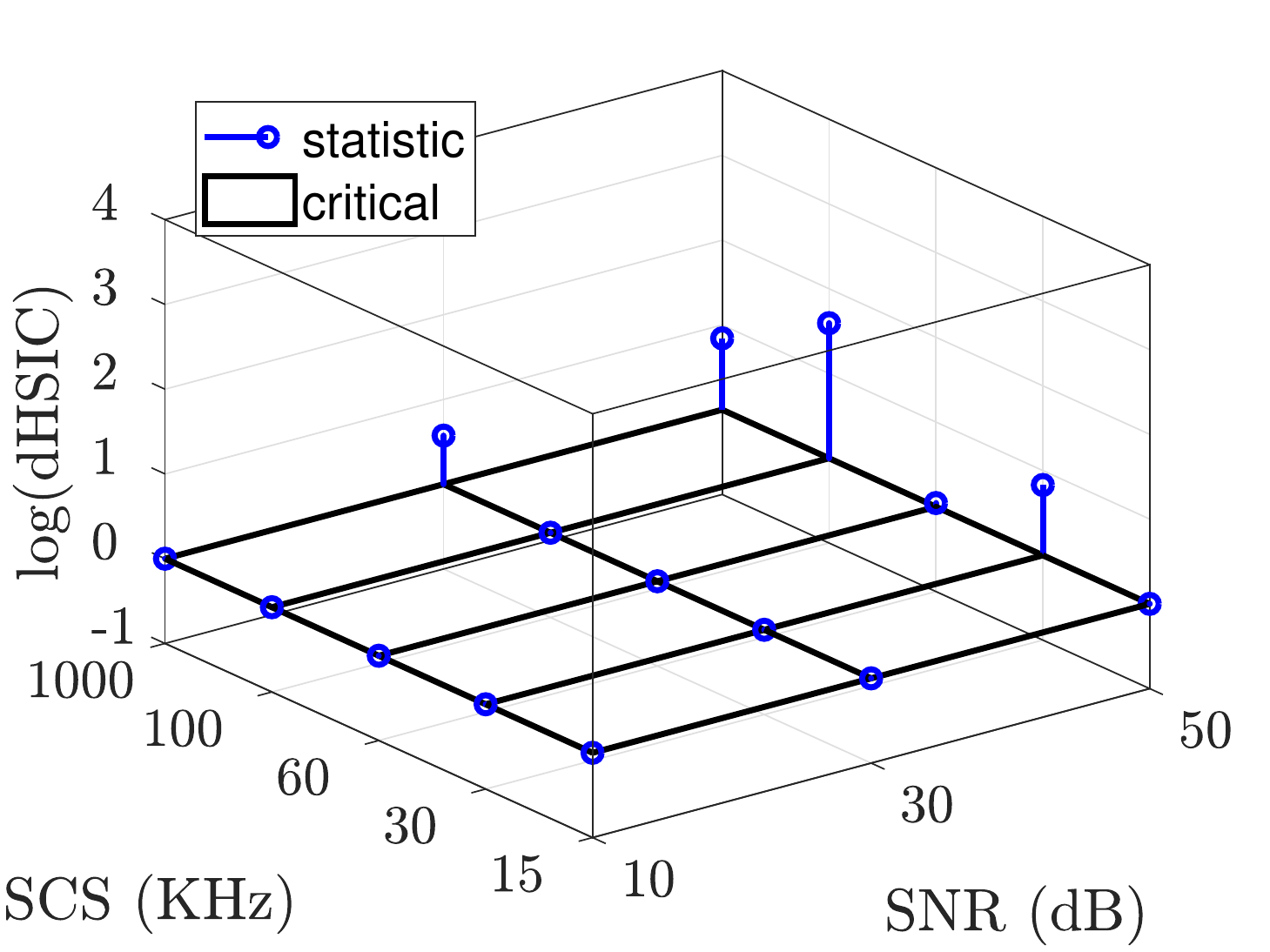}
         \caption{Residual for $\hat D=1$}
         \label{fig:SCC15dB2dim}
     \end{subfigure}
     \hfill
     \begin{subfigure}{0.32\textwidth}
         \centering
         \includegraphics[width=\textwidth]{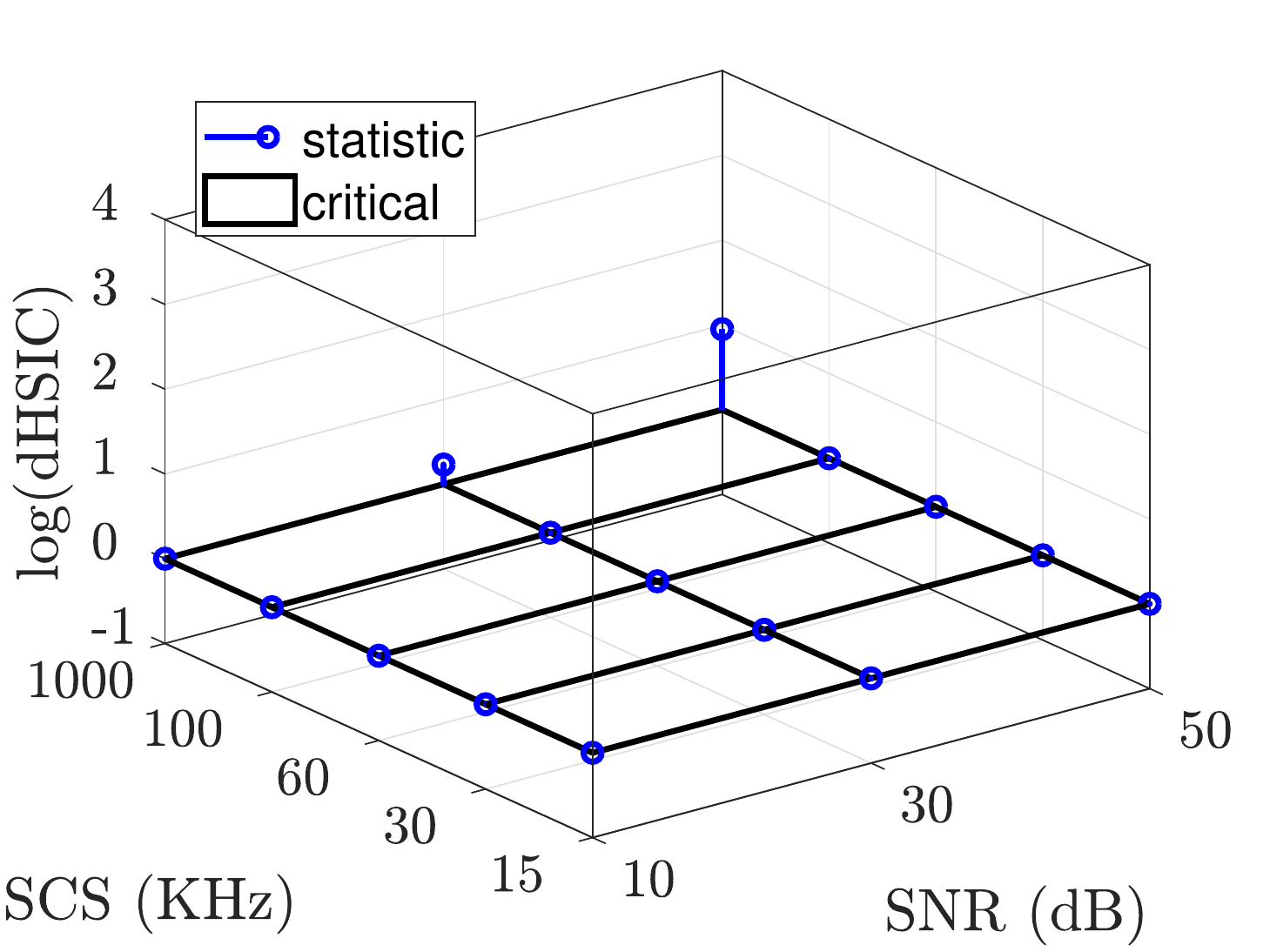}
         \caption{Residual for $\hat D=2$}
         \label{fig:SCC15dB4dim}
     \end{subfigure}
     \vfill
     \begin{subfigure}{0.32\textwidth}
         \centering
         \includegraphics[width=\textwidth]{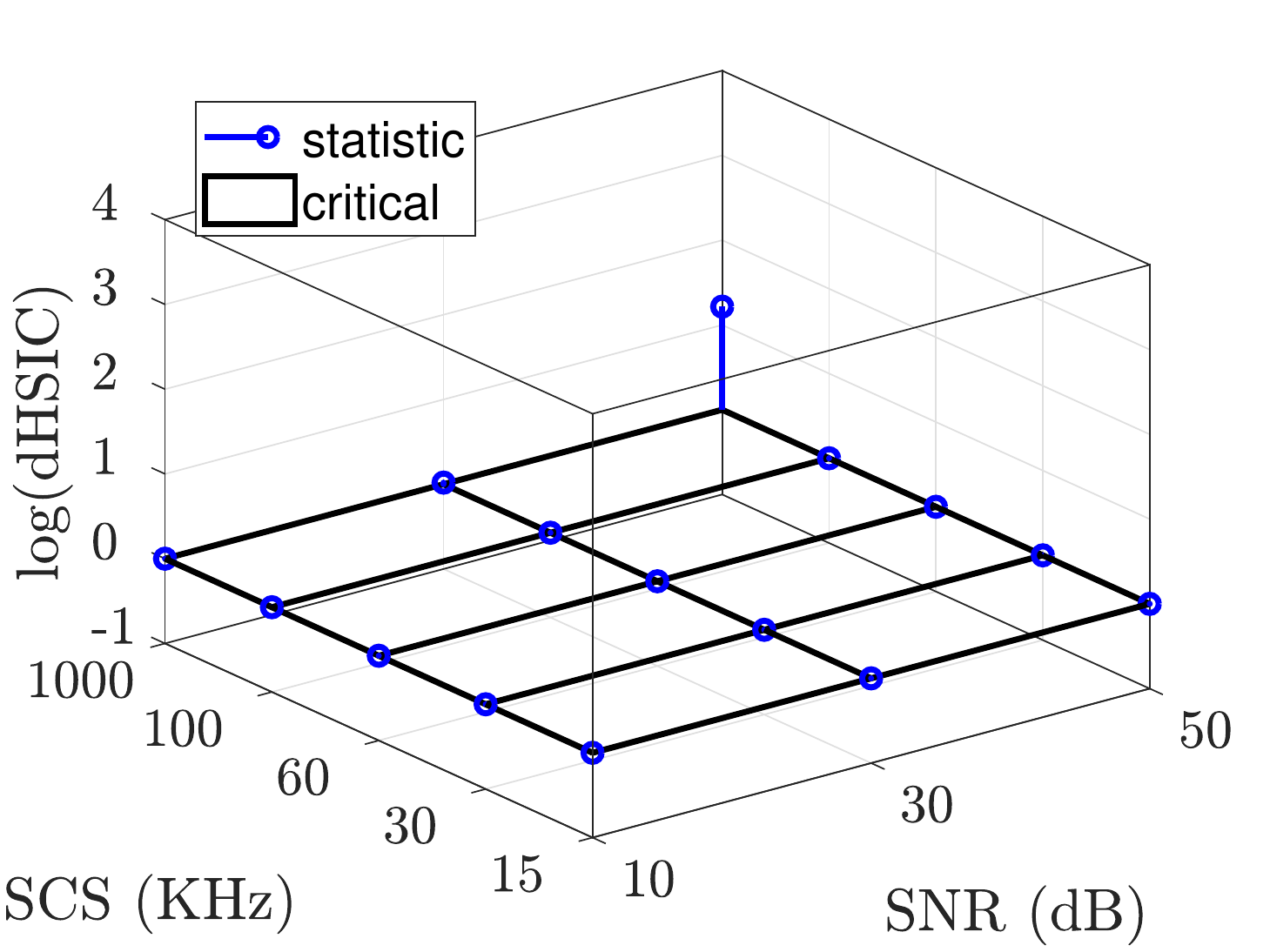}
         \caption{Residual for $\hat D=3$}
         \label{fig:SCC15dB4dim}
     \end{subfigure}
     \hfill
     \begin{subfigure}{0.32\textwidth}
         \centering
         \includegraphics[width=\textwidth]{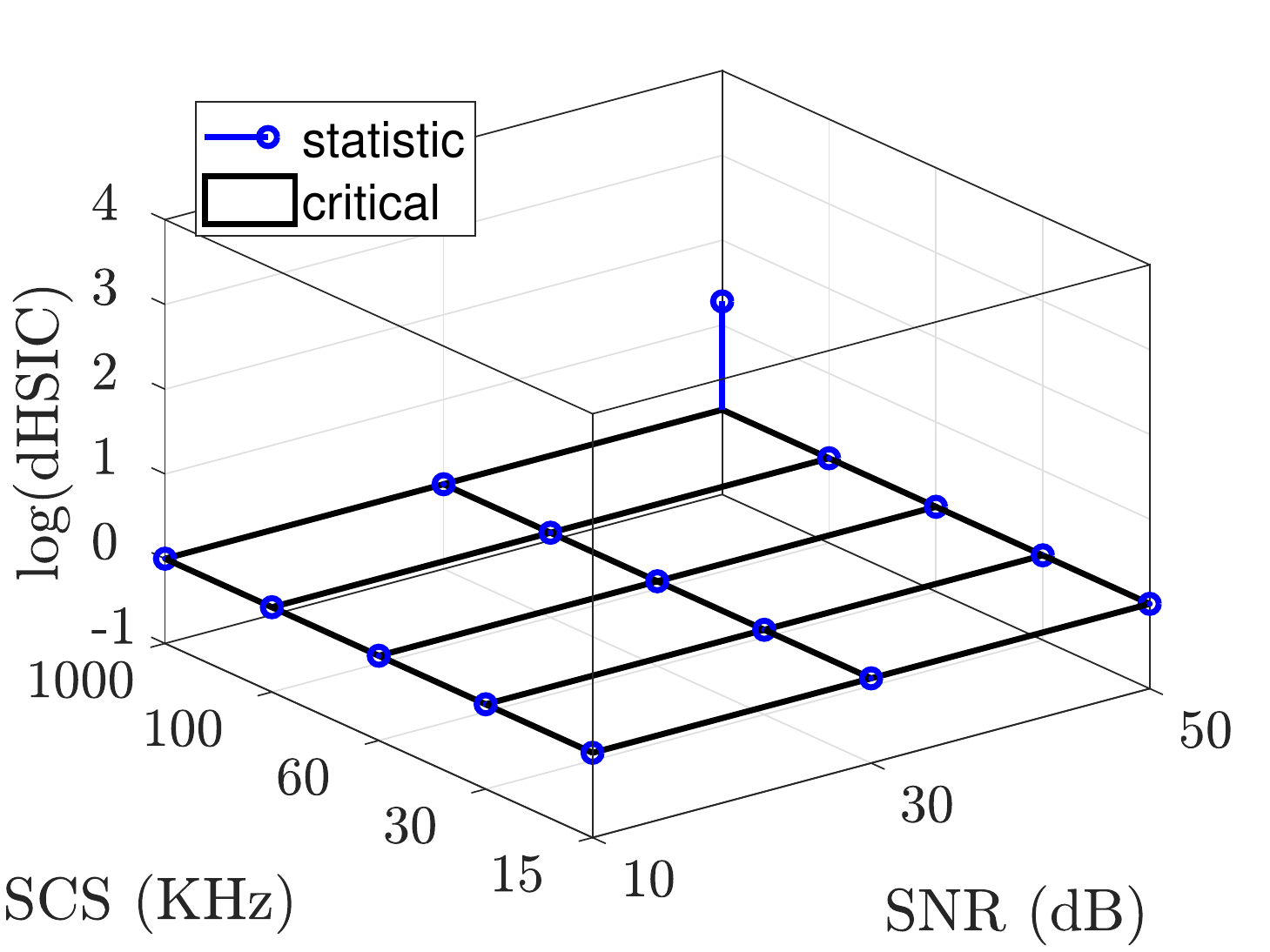}
         \caption{Residual for $\hat D=4$}
         \label{fig:SCC15dB4dim}
     \end{subfigure}
     \hfill
     \begin{subfigure}{0.32\textwidth}
         \centering
         \includegraphics[width=\textwidth]{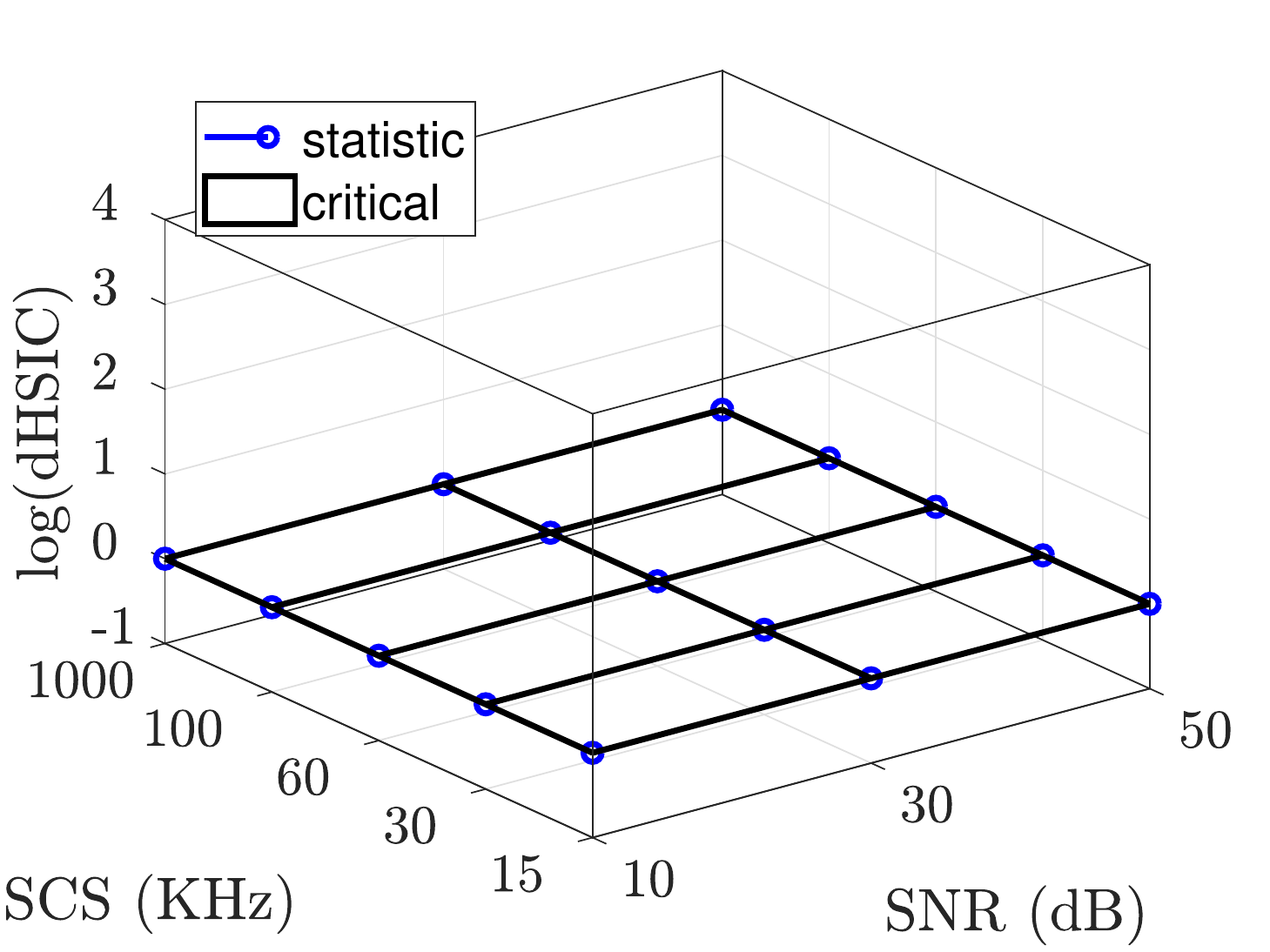}
         \caption{Residual for $\hat D=8$}
         \label{fig:SCC15dB4dim}
     \end{subfigure}
    \caption{$\log(dHSIC)$ of the test statistic and the critical values across the 32 subcarriers, for  $SCS = \{15, 30, 60, 100, 1000\} \text{ KHz}$ and $SNR = \{10,30,50\} \text{ dB}$.}
        \label{fig:SCC}
\end{figure*}

\subsection{Independence Criterion}
Note that the Pearson correlation coefficient assumes the presence of a linear relationship between the residuals. Furthermore, low values of the correlation coefficient do not necessarily mean low non-linear relationship between the residuals. Therefore, a second approach that considers the non-linear forms of dependence between the residuals is crucial. 
In this sub-section we explore $dHSIC$, an $M$-dimensional criterion to distinguish if the multivariate random variables (RV) are mutually independent \cite{dhsic}. The test applies a positive-definite kernel on the $M$-dimensional RV and maps its distribution into the reproducing kernel Hilbert space. The RVs are mutually independent if and only if the square distance of the transformed RVs is (close to) zero. More precisely,  $\mathbf{Z}=\left(\mathbf{z}^{1},\cdots,\mathbf{z}^{N}\right)$ be an $M \times N$ matrix based on the observations of the $M$-dimensional $\mathbf z^{i} =[z^i_1,\cdots,z^i_M]^T$ for $i\in[1,\cdots,N]$. The null hypothesis indicates that the $\mathbf{z}^{i}$ for  $i\in[1,\cdots,N]$ are mutually independent,
\begin{equation}\label{null}
H_0: \quad {F_{\mathbf{z}^1,\cdots,\mathbf{z}^N}}=F_{\mathbf{z}^1}{\cdots}F_{\mathbf{z}^N}
\end{equation}
whereas the alternative
\begin{equation}\label{alternative}
H_A: \quad {F_{\mathbf{z}^1,\cdots,\mathbf{z}^N}}\neq F_{\mathbf{z}^1}{\cdots}F_{\mathbf{z}^N}
\end{equation}
denotes that $\mathbf{Z}$ consists of at least two dependent vectors. An estimator $\widehat{dHSIC}_M$ of the statistical functional is as follows \cite[Def 2.6]{dhsic}:
\begin{equation}\label{dhsicest}
\begin{aligned}
     \widehat{dHSIC}_M(\mathbf{Z}) &= \frac{1}{M^2}{\sum_{i,j=1}^{M}\left( \prod_{l=1}^{N}\left(1{\ast}\left(K_{ij}^{l}\right)  \right)\right)} \\ 
    & + {\frac{1}{M^{2N}}}{\prod_{l=1}^{N}\left(\sum_{i,j=1}^{M} \left(K_{ij}^{l}\right)\right)} \\
    &- {\frac{1}{M^{N+1}}}{\sum_{i,j=1}^{M}}\left(\prod_{l=1}^{N}\left(1{\ast}\mathbf{1}_M \left( K_{ij}^{l}\right)\right) \right),
\end{aligned}
\end{equation}
where the operator $\ast$ denotes the Hadamard product and $\mathbf{1}_M$ is an $M \times 1$ vector of ones. Also, $\mathbf{K}^{l}=\left({\mathbf{K}^{l}_{ij}}\right)=\left(k^{l}(x_i,x_j)\right)\in\mathbb{R}^{M\times{M}}$ is the Gram matrix of the positive semi-definite Gaussian kernel $k^{l}$, defined $\forall{x_i,x_j}\in\mathbb{R}$ by, $$k^{l}=\exp\left(-\frac{\norm{x_i-x_j}^2}{\sigma^2}\right),$$
with bandwidth $\sigma=\sqrt{\frac{\text{med}\left({\norm{x_i-x_j}^2} \right)}{2}}$, where $\text{med}(.)$ is the median heuristic. Under the null hypothesis the asymptotic behaviour of $\widehat{dHSIC}_M$ is given by \cite[Theorem 3.1]{dhsic},
\begin{equation}\label{dhsicasym}
\small{
    N\widehat{dHSIC}_M \xrightarrow[]{d}\binom{2N}{2}\sum_{j=1}^{\infty}\alpha_{i}W_{i}^{2}, \quad N\xrightarrow{}\infty,
    }
\end{equation}
where $\xrightarrow[]{d}$ denotes convergence in distribution, $(a_i:i\in\mathbb{N})$ are constants and $\left(W_i:i\in\mathbb{N}\right)$ is an independent and identically distributed sequence of standard normal random variables.
Hence, to obtain the critical values of $H_0$ we need to approximate equation (\ref{dhsicasym}). Based on \cite{dhsic}, we implement the permutation test by constructing the re-sampling distribution function $\widehat{dHSIC}\left(\tilde{\mathbf{Z}}\right)$, where $\tilde{\mathbf{Z}}=\left(r_1(\mathbf{z}^{1}),\cdots,r_N(\mathbf{z}^{N})\right)$, where $r_1, \cdots, r_N$ are random resamplings without replacement. Finally, the critical value $CV$ for a specific significance level $\alpha$ is given by the $q$th element of the vector $\mathbf{D}^{HSIC}$ that contains the $B$ Monte-carlo realisations of $\widehat{dHSIC}{(\tilde{\mathbf{Z}}})$ in the increasing order. In other words, $CV_{\alpha}=\left[\mathbf{D}^{HSIC}\right]_{q},$ and
\begin{equation}
\small{
    q = 
    \ceil*{(B+1)(1-\alpha)}+\sum_{i=1}^{B}{  \mathds{1}_{\{\widehat{dHSIC}(\mathbf{Z})=\widehat{dHSIC}(\tilde{\mathbf{Z}}_i)\}} }
    }
\end{equation}
if $q \leqslant{B}$ and $\infty$ otherwise. The  operator $ \ceil*{.}$ denotes the ceiling function.

\section{Experimental Results}

To perform simulations, we obtain the CSI from $N= 400$ equi-distant (say $10~m$) spatial locations within a square area on the ground, between $x=100$ and $x=290$ and $y=-100$ and $y=90$ and the receiver at the location $(x,y,z) = (0,0,10)$ for $M=32$ sub-carriers, starting from a frequency of $2$~GHz, a sub-carrier spacing (SCS) of $SCS = \{15,30,60,100,1000\}$ KHz,  using the popular Quadriga channel models \cite{jaeckel2014quadriga}. 

First, we illustrate the effect of pre-processing, for a channel with $SCS = 100$ KHz and $SNR = 50$ dB. Fig. \ref{fig: OPR} (a) represents the magnitude of the observed channel, and  Fig. \ref{fig: OPR} (b) represents the magnitude of the predictable part across the $400$ locations for all the $32$ subcarriers. Note that the removal of the predictable part of the channel, which is shown to model the observed channel faithfully, results in the apparent independence of the residuals in Fig. \ref{fig: OPR} (c).

In Table \ref{Table:Corr}, we determine the average correlation coefficient of the residuals across the locations in a neighbourhood for $SNR=\{10,30,50\}$ dB and $\widehat{D} = \{1,2,3,4,8\}$. Note that, across all values of SCS and SNR, the correlation coefficient decreases with the removal of more dominant components $\hat{D}$. However, the decrease is less pronounced in SCS $\geq 100KHz$ because of the frequency selective nature of the channel response across the subcarriers. Therefore, more dimensions are required to characterize and remove the predictable part of the CSI. On the other hand, with an increase in the noise power, the correlation coefficient decreases. However, this may lead to an increase in the mismatch probability between legitimate users. In order to ensure that the residual sequence is completely  unpredictable, we also check if $\left\{\hat {\bm{z}}_{nu}(\widehat{D})\right\}_{n=1}^{400}$ are independent across the subcarriers.
In Table \ref{Table:Ind}, we tabulate the rejection rate of the null hypothesis of independence applying the HSIC across the locations in a neighbourhood. Like the mean correlation coefficient, the rejection rate decreases with a decrease in the SCS and SNR. 

To verify the independence between the $32-$ subcarriers at each of the locations, we depict the average critical value and the average test statistic value of the $dHSIC$ in Fig. \ref{fig:SCC}.  Similar to the case of spatial decorrelation and independence of locations in Tables I and II, the features also get disentangled from each other with an increase in $\widehat{D}$. The decrease is more pronounced as the number of $\widehat{D}$ increases. More precisely, even for $\widehat{D}=1$ the dependency between the subcarriers is strong only for $SCS=60$ dB and specific values of $SNR$, while for $\widehat{D}\geqslant{3}$ strong dependence exists only for $SCS=1000$ and $SNR=50$ dB. Furthermore, as shown in Fig. \ref{fig: OPR}, there is a strong dependence among the subcarriers for the observed channel for $SNR\geqslant{30}$ and all chosen $CSI$ values. Therefore, the implementation of the PCA-based pre-processing step seems to decrease the dependency between the subcarriers efficiently.        

\section{Conclusions and Future Work}
In this paper, we have shown that it is possible to employ unsupervised learning to separate the predictable and unpredictable components of the CSI to perform RF fingerprinting and SKG. In practice, the first two or three principal PCA components suffice to capture most of the predictable part of the CSI. 
In future work, we propose to estimate the amplitudes using ARMA modelling \cite{yuan2020machine,wu2021channel} that could separate the deterministic and the non-deterministic part of a zero-mean covariance stationary process as an approximation of the Wold representation theorem. Also, we propose to use other ML techniques like autoencoders with input layers of convolutional neural networks and long short-term memory networks to capture the spatio-temporal dynamics. 

\end{document}